\documentclass[aps,prb,twocolumn,superscriptaddress,floatfix,a4]{revtex4}
\usepackage{epsfig,amsmath,amssymb,color}
\usepackage{graphicx,times,epsfig,color}
\begin{document}

\title{
Real-time simulations of nonequilibrium transport in the single-impurity Anderson model}
%%%%%%%%%%%%
%%% AUTHORS
%
%
\author{F. Heidrich-Meisner}
\affiliation{Institut f\"ur Theoretische Physik C, RWTH Aachen University,  52056 Aachen, Germany, 
and\\ JARA - J\"ulich Aachen Research Alliance, Forschungszentrum J\"ulich, Germany}
\author{A. E. Feiguin} 
\affiliation{Microsoft Project Q, University of California, Santa Barbara, CA 93106, USA} 
\affiliation{Condensed Matter Theory Center, University of Maryland, College Park, MD 20742, USA}
\author{E. Dagotto}
\affiliation{Materials  Science and Technology Division, Oak Ridge National Laboratory,
 Oak Ridge, TN 37831, USA and\\
 Department of Physics and Astronomy, University of Tennessee, Knoxville,
 TN 37996, USA}

%%%%%%%%%%%%%%%%%%%%%%%%%%%%%%%%%%%%%%%%%%%%%%%%%%%%%%%%%%%%%

\begin{abstract}

One of the main open problems in the field of transport in 
strongly interacting nanostructures is the understanding of currents 
beyond the linear response regime. In this work, we consider  the 
single-impurity Anderson model and use the adaptive time-dependent density matrix
renormalization group (tDMRG) method to compute real-time currents out of equilibrium.
We first focus on the particle-hole symmetric point where Kondo correlations 
are the strongest and then extend the study of the nonequilibrium transport to the 
mixed-valence regime. As a main result, we present  accurate data for the current-voltage
characteristics of this model.
\end{abstract}
\date{June 16, 2009}
\maketitle

%%%%%%%%%%%%%%%%%%%%%%%%%%%%%%%%%%%%%%%%%%%%%%%%%%%%%%%%%%%%%%%%%%%%%%%%%%%%%%
%%%%%%%%%%%%%%%%%%%%%%%%%%%%%%%%%%%%%%%%%%%%%%%%%%%%%%%%%%%%%%%%%%%%%%%%%%%%%%
%%%%  SECTI

\section{Introduction}

Experiments in strongly interacting nanostructures are 
often well described by relatively simple model
Hamiltonians. In several cases, these models are integrable 
and can be solved exactly, \cite{andrei80} or they 
can be studied using powerful numerical methods, 
such as the numerical renormalization group (NRG).\cite{wilson75,bulla08}
These techniques have had enormous success in describing the equilibrium properties 
of these models. However, the understanding of open systems out of equilibrium 
remains an extremely challenging area of research.
 While recent studies have succeeded in 
calculating the current-voltage (I-V) characteristics of the interacting resonant
level model, \cite{mehta06,doyon07,borda07,boulat08,nishino09} the focus of current research has been devoted toward the more difficult single-impurity
Anderson model incorporating Kondo correlations.
For example, experimental results for the finite-bias transport properties of a quantum
dot still await a complete theoretical explanation.\cite{goldhabergordon98,wiel00,grobis08,scott09}
Besides the fundamental interest in solving this type of problems,
Kondo physics is an emergent feature in many experiments on nanoscopic 
systems.\cite{goldhabergordon98,wiel00,grobis08}

Our goal here is to calculate the I-V characteristics of the single-impurity 
Anderson model, described
by the Hamiltonian $H=H_{\mathrm{dot}}+H_{\mathrm{leads}}$:\cite{hewson}
\begin{eqnarray}
H_{{\mathrm{dot}}} &=& V_{\rm g} \sum_{\sigma} n_{1\sigma} + U/2 \sum_{\sigma}
n_{1\sigma} n_{1\bar{\sigma}} \nonumber\\ 
H_{\mathrm{leads}} &=& - t^{\prime} \sum_{l=R,L;\sigma} (c_{1,\sigma}^{\dagger} c_{l,1,\sigma}+h.c.)\nonumber \\
       && -\sum_ {n=1;l=R,L;\sigma}^{N_{L,R}} t_n\,(c_{l,n\sigma}^{\dagger} c_{l,n+1 \sigma} ~+~
h.c. )\,.\label{eq:ham}
\end{eqnarray}
 $H_{\mathrm{dot}}$ contains the gate potential
$V_{\rm g}$ and the Coulomb repulsion $U$ acting on the dot. The first term
in $H_{\mathrm{leads}}$
is the hybridization of the dot with the conduction band, and the last term
describes the leads. $N_{L(R)}$ is the number of sites 
on the left (right) lead, with $N=N_L+N_R+1$. 
$c_{\nu,\sigma}^{\dagger}$ is a fermionic creation operator acting on the dot ($\nu=1$) 
or a site $i$ in the left (right) lead ($\nu=L(R),i$),
with a spin index $\sigma=\uparrow,\downarrow$.
      
The Kondo resonance is a signature of the hybridization of the leads with a single electron on
the dot.\cite{hewson} Electrons from the leads screen the extra spin, 
forming an overall singlet, and the
corresponding Kondo cloud.\cite{hewson} This collective state can be suppressed, and eventually destroyed, by temperature, magnetic field, or a large bias between the leads. The ultimate goal is to understand the subtle mechanisms behind the suppression of the Kondo resonance, and how this response translates into the I-V characteristics of the system. 
In particular, when a finite voltage bias is applied, 
results for the density of states (DOS) at the particle-hole
symmetric point $V_{\rm g}=-U/2$ seem to indicate \cite{meir93,wingreen94,fujii03,han07,anders08} that the Kondo peak smoothly
decreases, while the spectral weight is transferred to satellite peaks centered at the chemical potentials
of the leads at frequencies $\omega\approx \pm V/2$. At large biases, 
these peaks merge with  broad shoulders at $\omega\approx\pm U/2$.\cite{meir93,wingreen94,fujii03,han07,anders08}
 The possible existence of a negative
differential conductance in the high-voltage  regime of nanostructures is a topic of intense discussion.\cite{boulat08}

Beyond perturbative limits in $U$ or bias voltage $V$,
the properties of the steady state, including average currents, are still under scrutiny.
Therefore, to gauge the validity
of approximate analytical schemes used to study nonequilibrium transport,
it is highly desirable to have accurate numerical results available as a benchmark.
Several efforts to work out the nonlinear transport regime of this
model have been made using  perturbative\cite{fujii03,hershfield91,hershfield92} as well as renormalization schemes,
\cite{schoeller00,rosch01,doyon06,gezzi07,jakobs07,jakobs09} the non-crossing approximation,\cite{meir93,wingreen94} flow-equations,\cite{kehrein05} real-time path integral approaches,\cite{weiss08} Quantum Monte Carlo
(QMC),\cite{han07,werner09,schiro09}  and the Bethe ansatz.\cite{konik01} 
 Recently, the time-dependent 
NRG method has been used to simulate 
aspects of nonequilibrium physics in quantum dots.\cite{anders05,anders06,roosen08,anders08}
In this work, we introduce the tDMRG \cite{white04,daley04} to address 
two questions: first, what is the
I-V characteristics at $V_{\rm g}=-U/2$ and, second, how do nonequilibrium
currents behave in the  mixed-valence regime. 
Furthermore, tDMRG gives  access to the transient regime, {\it i.e.}, information on how the steady state is reached,
which some of the aforementioned methods are, as of now, not designed for.
We will also
provide a technical discussion of our tDMRG calculations, in terms of
system sizes and accuracy needed to obtain high quality results.
We work at
values of $U/\Gamma\leq 8$, where $\Gamma =2\pi \rho_{\mathrm{leads}}(E_F)t'^2 =2t'^2$ is the
hybridization parameter, with $\rho_{\mathrm{leads}}(E_F)$ the density of states in the tight-binding leads at the Fermi energy $E_F$.
Our approach employs a real-space
 representation of the leads with $t_n=1$, which we find is a useful modeling of the leads for bias values 
 $V \gtrsim T_{\rm K}$, where $T_{\rm K}$ is the Kondo temperature.
In the case of $V_{\rm g}=-U/2$, our results are in excellent agreement with those from the
functional renormalization group (fRG)\cite{jakobs09}
for $U/\Gamma\leq 6$.

The plan of this exposition is the following. In Sec.~\ref{sec:dmrg},
we shall introduce our numerical approach, the tDMRG, and we 
discuss the error sources and the quality of our numerical data.
In Sec.~\ref{sec:ph}, we present our results for the particle hole-symmetric point, while
in Sec.~\ref{sec:mvregime}, we turn to the mixed-valence regime. 
We conclude with a summary and discussion of our results in Sec.~\ref{sec:sum}.

\section{Numerical Method: DMRG}
\label{sec:dmrg}

 The tDMRG method has previously been used 
 in Refs.~\onlinecite{alhassanieh06} and \onlinecite{dasilva08}
to study the linear conductance of
the single-impurity Anderson model Eq.~\eqref{eq:ham}.
In the latter
effort, it has been argued that a logarithmic
discretization,\cite{bulla08} {\it i.e.}, $t_n\propto
\lambda^{-n/2}$ is useful in capturing Kondo correlations in the linear conductance at $V_{\rm g}=-U/2$ and 
using this trick,  perfect conductance was obtained for
$U/\Gamma\lesssim 7$. Other applications of the tDMRG to transport in nanostructures
include the (interacting) resonant level model,\cite{guo09,boulat08} short chains of spinless
fermions\cite{schneider06,schmitteckert04} as well as nonequilibrium transport through a point contact in a non-Abelian 
quantum Hall state.\cite{feiguin08c} 
Moreover, first tDMRG results for Eq.~\eqref{eq:ham}
out of equilibrium  and at $V_{\rm g}=-U/2$ 
were presented in Ref.~\onlinecite{kirino08} by Kirino {\it et al.} for $U/\Gamma\leq
2.8$ and in Ref.~\onlinecite{dasilva08} for $U/\Gamma=3.125$. We here move 
substantially beyond the
scope of these previous studies\cite{kirino08,dasilva08} as we consider 
larger values of $U/\Gamma$, perform a 
finite-size scaling analysis, and use up to $m=2000$ DMRG states during the time-evolution (compared to $m=800$ in Ref.~\onlinecite{kirino08}).

\begin{figure}[t]
\centerline{\includegraphics[width=0.45\textwidth]{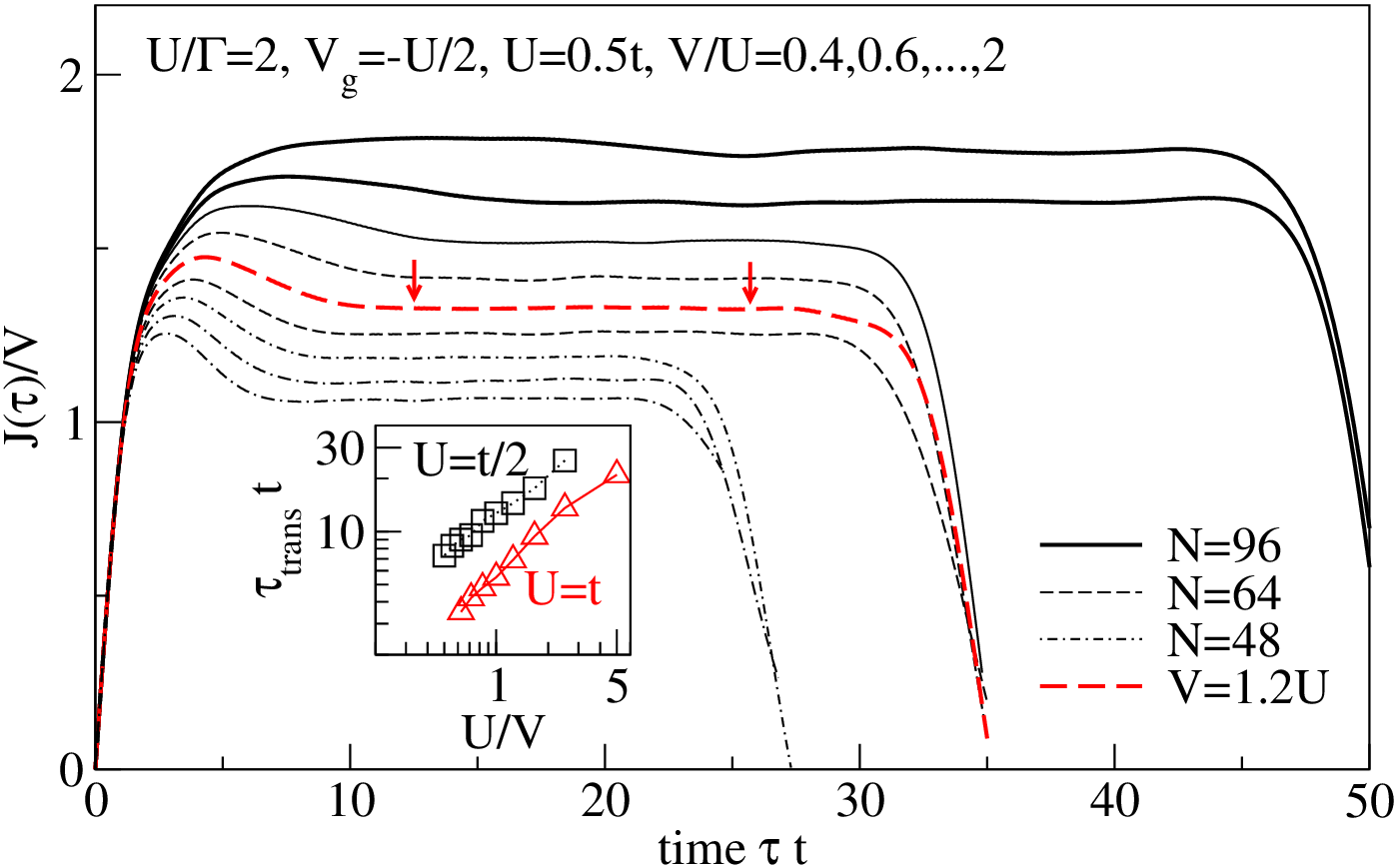}}
\caption{(Color online) Particle-hole symmetric point, $V_g=-U/2$: Current $J(\tau)/V$ vs. time $\tau$ for $U/\Gamma=2$ ($U=0.5t$, $t'=0.3535t$,
$V_{\rm g}=-U/2$) for  $V/U=0.4,0.6,\dots,2$ [$N=96$ (thick lines);  $N=64$ (thin lines); $N=48$ (dot-dashed lines)]. 
The arrows indicate the time-interval used for the calculation of  the steady-state current
$J_{\mathrm{steady}}=\langle J(\tau)\rangle_{\tau} $ for $V/U=1.2$ from a
time average $\langle \, \cdot \,\rangle_{\tau} $.\cite{dasilva08} 
 Inset: estimates of the transient time $\tau_{\mathrm{trans}}$ as a function of bias for $U=t/2$ (squares, $\Gamma=t/4$) and $U/t=1$ (triangles up, $\Gamma=t/2$), both at $U/\Gamma=2$.}\label{fig:crr}
\end{figure}

In our numerical calculations,\cite{alhassanieh06,dasilva08} we first compute the ground state of
Eq.~\eqref{eq:ham} and then, at time $\tau=0^+$, we apply an extra term $H_{\mathrm{bias}}$ and evolve under
$H+H_{\mathrm{bias}}$:
\begin{equation}
H_{\mathrm{bias}}=\frac{V}{2}\sum_{n=1}^{N_R}
{n}_{R,n} -      \frac{V}{2}\sum_{n=1}^{N_L}
{n}_{L,n}\,.
\end{equation}
In principle, different spatial forms of 
$V$ can be used,\cite{white04,schneider06} 
which sometimes affect the transient behavior but leave  the steady-state
currents  unchanged.
 We measure the expectation value $J(\tau) $ of the local current operator acting
on the links connecting the dot to the leads, averaging over the right and left
links.\cite{alhassanieh06,dasilva08} We set
$e=\hbar=1$. All simulations are performed at an overall half filling of dot and leads.

In our tDMRG runs, we use a Trotter-Suzuki time evolution scheme with time-steps 
of $0.05\leq \delta \tau \,t \leq 0.15$. We find a fast convergence with
system size
on chains with $N_L(N_R)$ odd(even).\cite{alhassanieh06,hm09}
Our goal is to achieve a numerical  accuracy of $\delta J_{\mathrm{steady}}/V \sim 0.01 G_0$,
where $G_0$ is the conductance quantum. The steady-state currents are computed by taking suitable 
time-averages, {\it i.e.}, $J_{\mathrm{steady}}=\langle J\rangle_{\tau} $.  In principle, the true steady state is only reached for $N\to \infty$,
while on finite systems, the term quasi-steady states may be more appropriate.
For simplicity, we do not make this distinction here.

The error sources in tDMRG are:\cite{gobert05,dasilva08}
(i) truncation errors, (ii) the Trotter error, and (iii) the determination of 
the time intervals to extract average currents from the real-time data at a given bias. Typically, we perform
 runs with different discarded weights $\delta \rho$ at fixed $U, \Gamma, V$ and $N$ to
identify the time windows for the averaging.
We find (i) and (iii) to be the dominant sources in the present problem as the Trotter errors can
effectively be reduced by using small time-steps and this error grows only linearly with the simulation time.\cite{schollwoeck05}
The truncation errors can be suppressed by
decreasing the discarded weight  $\delta \rho$ (Ref.~\onlinecite{schollwoeck05}) 
during the time evolution. This error source is thus well-controlled, too.
As for the averaging procedure, the determination of steady-state regimes are the 
least unambiguous for large gate voltages, intermediate biases voltages $V\gtrsim T_{\rm K}$,
 and/or large $U/\Gamma \gtrsim 6$. There, the accuracy
of our data for $\langle J \rangle_{\tau}/V$ is worse than $0.01 G_0$. 

\begin{figure}[t] 
\centerline{\includegraphics[width=0.45\textwidth]{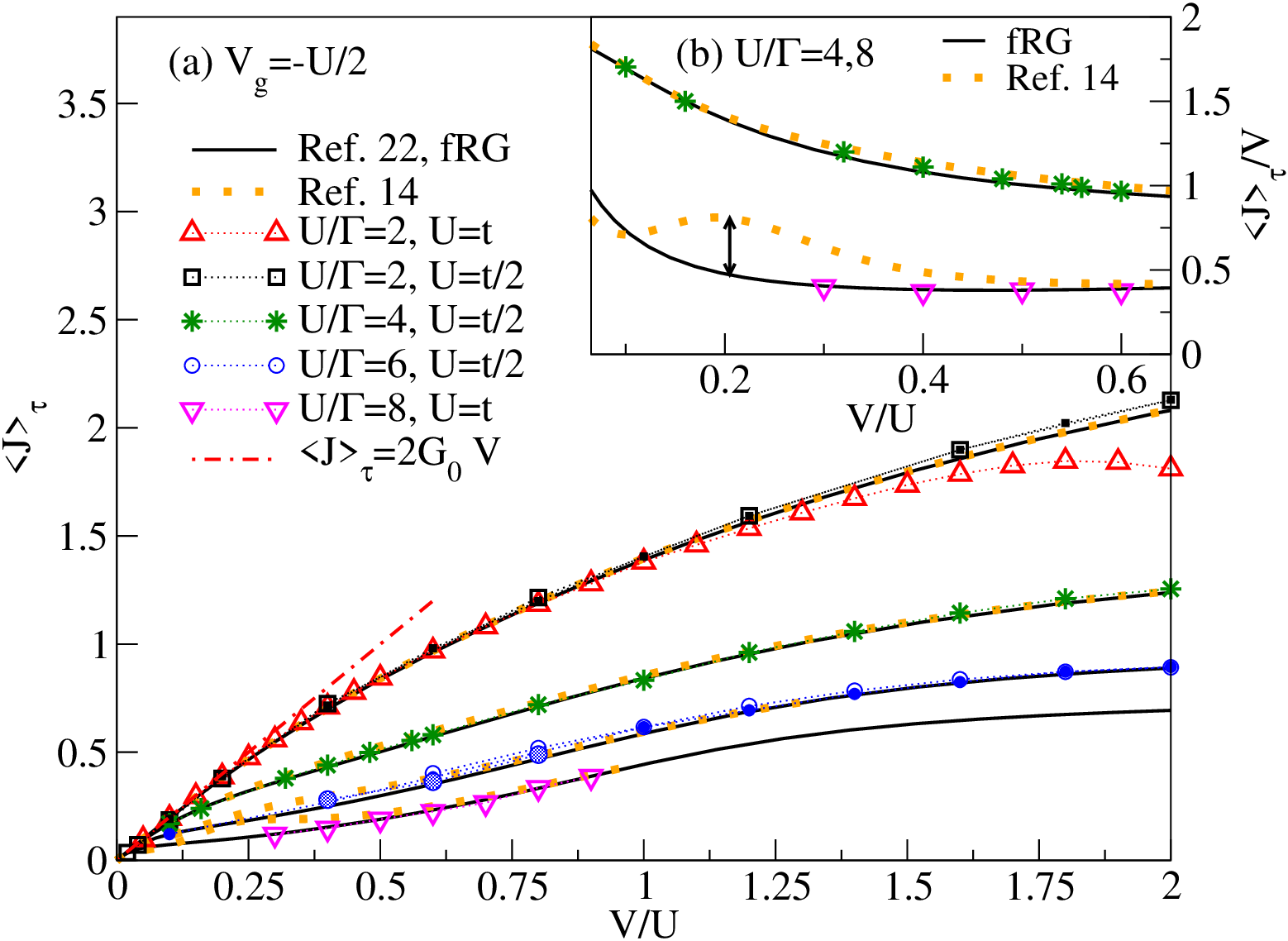}}
\caption{(Color online) Particle-hole symmetric point, $V_g=-U/2$:
(a) Current-voltage curves for $U/\Gamma=2,4,6$ with $U=t/2$ (squares, stars, circles) as well
as  data for $U/\Gamma=2,8$ with $U=t$ (triangles down and up, respectively). 
Open symbols: $N=64$, full ones: $N=96$ sites, shaded ones: $N=128$, stars and triangles down: finite-size extrapolations in $1/N$ ($U/\Gamma=4,8$ only).
Dash-dotted line: $\langle J\rangle_{\tau}=2G_0 V$.
(b) Enlarged view of the low bias region for $U/\Gamma=4$ and $8$, $\langle J\rangle_{\tau}/V$ vs. $V$. Solid lines: fRG results from
Ref.~\onlinecite{jakobs09}; dotted lines: 4th-order perturbation theory results  from Ref.~\onlinecite{fujii03}.
(from top to bottom). For $U/\Gamma=8$, the arrow in the inset indicates the difference between the fRG and tDMRG results on the one hand and those
from   Ref.~\onlinecite{fujii03} on the other hand.
For NRG results for $T_{\rm K}$, see, {\it e.g.},  Fig.~6 in Ref.~\onlinecite{dasilva08}.
}\label{fig:iv}
\end{figure}

An important  
aspect of any simulation of time-evolution is the entanglement growth  (see, {\it e.g.}, Ref.~\onlinecite{calabrese07}). 
This is measured through the entanglement entropy $S_l=-\mbox{tr}(\rho_l \mathrm{ln} (\rho_l))$,
where $\rho_l$ is the reduced density matrix for a block of length $l$, as is naturally accessed in
DMRG.\cite{schollwoeck05} 
As our set-up realizes a so-called global quantum quench, $S_l$ can grow significantly as a function
of time. 
As a consequence, to keep the truncated weight below a given threshold, an increasing
number of states needs to be kept during the time evolution. 
For  $V \gtrsim 0.1t$, it is  imperative to work at a fixed $\delta \rho$ (as compared to a fixed $m$)
during the time evolution  to have firstly, an efficient simulation and secondly, a meaningful
control over the numerical accuracy by studying the scaling in $\delta \rho$. 
Typically,  
to meet our target accuracy at $V\leq 2t$,  a 
discarded weight of $\delta \rho \lesssim 10^{-7}$ is found to be appropriate.

%%%%%%%%%%%%%%%%%%%%%%%%%%%%%%%%%%%%%%%%%%%%%%%%%%%%%%%%%%%%%%%%%%%
% RESULTS
%%%%%%%%%%%%%%%%%%%%%%%%%%%%%%%%%%%%%%%%%%%%%%%%%%%%%%%%%%%%%%%%%%%

\section{Results for particle-hole symmetry}
\label{sec:ph}

 After these remarks on the method, we turn to the discussion of our tDMRG results. The time-dependent currents at $V_{\rm g}=-U/2$ are displayed in Fig.~\ref{fig:crr} for $U/\Gamma=2$
($U=0.5t$) and $V/U=0.4,0.6,\dots,2.0 $.  
 Since we plot $J(\tau)/V$, the currents
decrease as the bias increases.
An example for the time-interval used to compute   the steady-state
currents is indicated by arrows for the case of $V=1.2U$ in the figure. 
While a full analysis of the short-time
transients and its dependence on both $V$ and $\Gamma$ (see also Ref.~\onlinecite{schiller00}) 
will be presented elsewhere, note that the transient times,
{\it i.e.}, the time for the current to reach its steady-state plateau,
 {\it decreases}  as the bias increases, 
to, {\it e.g.}, about $\tau \sim 6/t$ at 
$V\gtrsim 2U$ and $U=t/2$. This is illustrated in the inset of Fig.~\ref{fig:crr}, where we  estimate
the transient time from the time that it takes  the initial maximum in $J(\tau)$ to decay. 
For this reason, and due to less severe finite-size effects, short chains
of $N \sim 24,32,48$ are often sufficient  at large biases.

\begin{figure}[t]
\centerline{\includegraphics[width=0.45\textwidth]{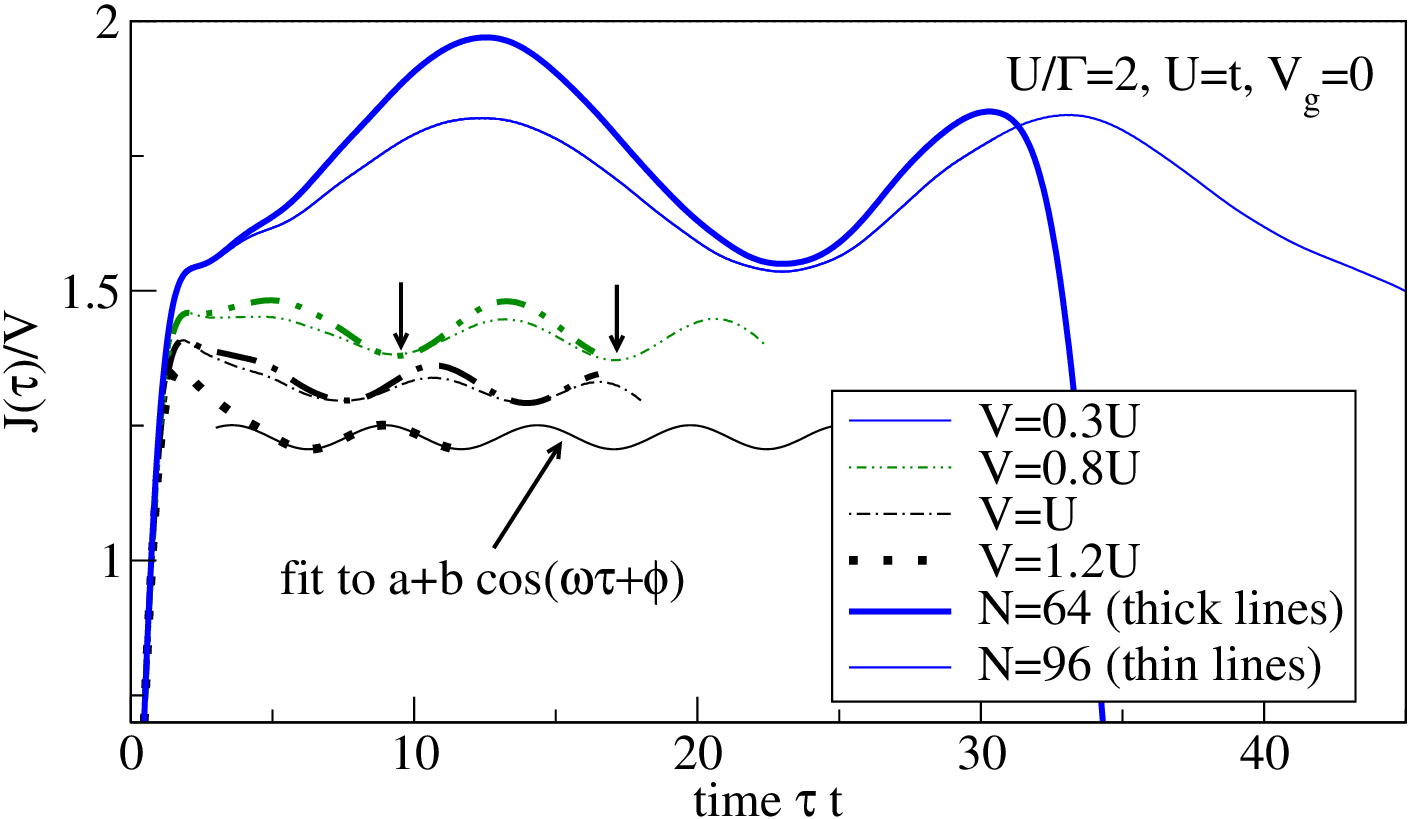}}
\caption{(Color online) Mixed-valence regime: Current $J(\tau)/V$ vs. time $\tau$ for $U/\Gamma=2$ ($U=t$, $t'=0.5t$,
$V_g=0$) and bias values $V/U=0.3,0.8,1, 1.2$ [$N=96$ (thin lines), $N=64$ (thick lines)].
Arrows indicate the interval used for the averaging for $\Delta
V=0.8U$ and $N=64$. The thin solid line for $V=1.2U$ is a fit of $a+b \cos(\omega\,\tau+\phi)$ to the corresponding $N=64$ 
tDMRG data
(dotted line).
}\label{fig:crr_vg}
\end{figure}

The average currents $\langle J\rangle_{\tau} $ are plotted as a function of bias in 
Fig.~\ref{fig:iv}(a) for $U/\Gamma=2,4,6$ ($U=0.5t$ for all three sets, plus  additional  
$U=t$ curves for $U/\Gamma=2$ and $8$).  
For $U/\Gamma = 2$ and $U=t$, the average current increases monotonically with bias and eventually
decreases. The latter gives rise to a negative differential conductance, yet this is simply
due to the decreasing overlap of the leads' DOS, which have a finite bandwidth of $4t$ in our
case. For that reason, namely the finite band-width, the current vanishes in the limit of $V\to \infty$ 
in our simulations. 

We can estimate at which voltage the band-curvature and the finite bandwidth, and
 thus the deviation from the
wide-band limit, start to play a role by varying $U$ while keeping $U/\Gamma$ fixed. 
Comparing $U=0.5t$ (squares) and $U=t$ (triangles up) results in Fig.~\ref{fig:iv}(a)
at $U/\Gamma=2$, we find that the smaller $U$s give the best agreement with fRG\cite{jakobs09} (thick, solid lines) at large voltages. Note that the fRG calculations at these small values of
$U/\Gamma\sim 2$ are expected to be very reliable.
 We thus mostly use $U=0.5t$ for our runs at larger values of $U/\Gamma$ as we  can then compare 
 to the fRG at larger values of $V$. This has the disadvantage that the transient time  
 increases with increasing $1/\Gamma$ (corresponding to a decreasing $U$ at a fixed $U/\Gamma; $see the inset of Fig.~\ref{fig:crr}), and therefore, for the runs at $U/\Gamma=8$ where we are interested in
 the behavior at intermediate bias values, we use $U=t$.

At $U/\Gamma=6$, a tendency of separating the narrow Kondo peak
 at small bias values from the Coulomb blockade peak at $V/U\gtrsim 0.5 $
 is visible. 
Unfortunately, at biases $V\sim 0.2U$, we observe a crossover in the monotonic behavior of
our finite-size data, with $\langle J\rangle_{\tau} $ increasing (decreasing) with system size at small (large) biases.\cite{dasilva08}
This renders this bias regime the most difficult for the determination 
of steady-state regimes. 

We emphasize the excellent agreement with the fRG results,\cite{jakobs09}
in particular,
for bias voltages $V>T_{\rm K}$. The differences between tDMRG and
fRG seen at larger biases for the $U=t$ curve are due to the wide-band limit taken in fRG with a flat DOS (linear dispersion), compared to 
a semi-elliptical DOS in tDMRG calculations, as discussed above. We further compare to the 4th-order 
perturbation theory results
(Ref.~\onlinecite{fujii03}; dotted lines in Fig.~\ref{fig:iv}). While overall, the three methods agree for $V>T_{\rm K}$ and $U/\Gamma=2$,  the perturbation theory
result lies slightly above both tDMRG and fRG for $U/\Gamma>4$ and in the interval $0.2U\lesssim V\lesssim 0.8 U$ (see Fig.~\ref{fig:iv}).  Note, though, that the
differences between fRG and Ref.~\onlinecite{fujii03} are, at $U/\Gamma=4$,  rather small
and difficult to resolve numerically, but become more significant at $U/\Gamma=6$ and 8.

Therefore, we also consider  the behavior for  $U/\Gamma=8$ and $ 0.3\lesssim V/U \lesssim 0.9$, for which the accuracy of our data is  reduced to $\delta \langle J \rangle_{\tau}/V \approx 0.03 G_0$,
due to an additional oscillatory component in the transient behavior that becomes more prominent at large $U/\Gamma$.\cite{schiller00} Nevertheless, as  Fig.~\ref{fig:iv}(b) clearly shows,
our tDMRG results are still in good agreement with fRG, while the corresponding 4th-order perturbation theory\cite{note1} predicts too high values of $\langle J \rangle_{\tau}/V$.
This constitutes an example where our numerical results serve to support the validity of one analytical approach, the fRG in this case, while 
unveiling that 4th-order perturbation theory\cite{fujii03} seems to break down at $U/\Gamma \gtrsim 6$ [see  Fig.~\ref{fig:iv}(b)].  Consequently, we may conclude that the double-minimum structure
that Ref.~\onlinecite{fujii03} predicts to exist in the differential conductance likely is an artifact of approximations. This double minimum is  precisely seen in the
bias window in which the differences between our tDMRG and the 4th-order perturbation theory are the largest.

\begin{figure}[t]
\centerline{\includegraphics[width=0.45\textwidth]{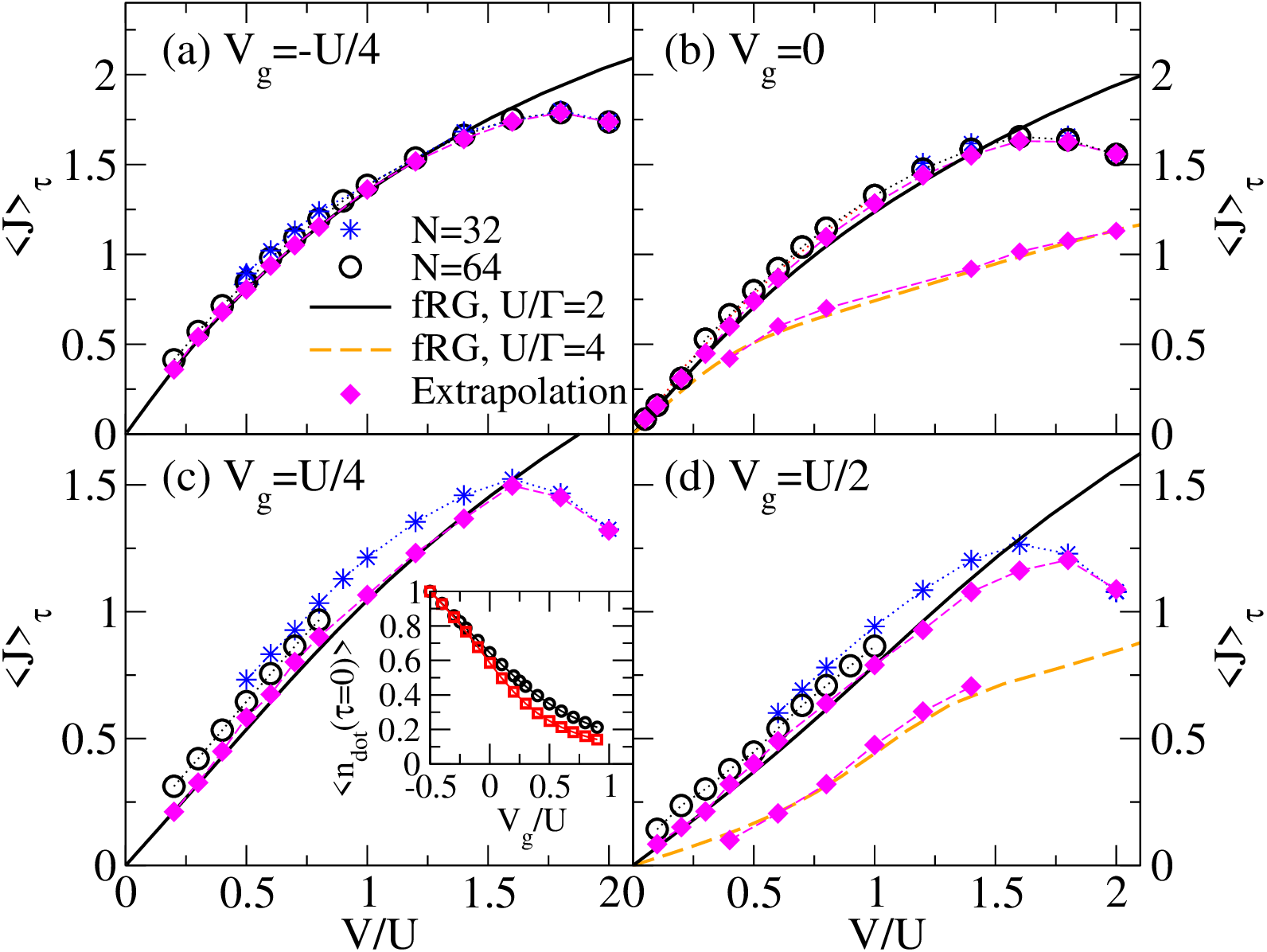}}
\caption{
(Color online) Mixed-valence regime: Current-voltage curves for $U/\Gamma=2$ ($U=t$,  symbols with dotted lines), 
$U/\Gamma=4$ [$U=t/2$, extrapolated data only, panels (b) and (d)], 
and several gate potentials: (a) $V_{\rm g}=-U/4$; (b) 
$V_{\rm g}=0$; (c) $V_{\rm g}=U/4$; and (d) $V_{\rm g}=U/2$.  
Solid diamonds 
are extrapolations 
of finite-size data in $1/N$.  
Inset in (c): $\langle n_{\mathrm{dot}}(\tau=0)\rangle$ vs. $V_{\rm g}/U$ for $U/\Gamma=2$ (circles) and 4 (squares).
}\label{fig:iv_mixed_2}
\end{figure}

As for the regime $V \lesssim T_{\rm K}$, note that 
at $V/U= 0.1$ and $U/\Gamma=2$, the $N=96$ result reaches about 
99 \% of the expected perfect conductance. 
At larger $U/\Gamma$ and 
$V\lesssim 0.2U$, the currents from $N\leq 128$ are below the fRG result.
Yet,  for $U/\Gamma=4$ and, {\it e.g.}, $V/U=0.1$ and 0.16, we have extrapolated
the finite-size data for  $\langle J\rangle_{\mathrm{\tau}}/V$ in $1/N$ 
[see Fig.~\ref{fig:iv}(b)]. The agreement with fRG is quite reasonable.

\section{Results for the mixed-valence regime}
\label{sec:mvregime}

We now turn our attention to the mixed-valence  regime $V_{\rm g}\not= -U/2$, concentrating on
$U/\Gamma=2$ and 4. Real-time currents are shown in Fig.~\ref{fig:crr_vg} for $U/\Gamma=2$ ($U=t$)
and $V_{\rm g}=0$.
In contrast to the $V_{\rm g}=-U/2$ case, 
significant oscillations with a period $T\propto 1/V$ (Ref.~\onlinecite{schiller00}) in the current are observed. 
These  are 
due to finite-size gaps and their amplitudes decay with $1/N$. 
In principle,
similar oscillations can be seen in the particle-hole symmetric case in the two local currents connecting the dot to the left and right lead.
As in computing $\langle J \rangle_{\tau} $, we average over these two local currents and as they oscillate with  a phase shift of $\pi$,\cite{schneider06}
the oscillations cancel out in $\langle J \rangle_{\tau} $ at $V_g=-U/2$, but not away from that special point.
For a detailed discussion of such oscillations in tDMRG calculations of real-time currents, see Ref.~\onlinecite{schneider06}.  
We exploit the existence of these oscillations to determine intervals to compute
steady-state currents from,\cite{schneider06} averaging over a full period (see the arrows in 
Fig.~\ref{fig:crr_vg} and the fit to the $V=1.2U$ data for an illustration).

The average currents $\langle J\rangle_{\tau}$ are plotted vs. bias in Fig.~\ref{fig:iv_mixed_2} 
for $U/\Gamma=2$ and $4$. For $U/\Gamma=2$, we include the results from
system sizes of $N=32$ and $N=64$  sites to illustrate the emergent finite-size effects.
To our advantage, we find that at large biases, even small chains of 32 sites yield
well-converged results, as the example of $V_{\rm g}=-U/4$ 
and $U/\Gamma=2$ shows 
[see Fig.~\ref{fig:iv_mixed_2}(a)]. Although we have not pursued this, it is reasonable to expect that at bias values
of $V/U\gtrsim 2$, exact diagonalization may produce useful results as well.

For several values of the bias, we have extrapolated the average currents in $1/N$ (shown as solid
diamonds in Fig.~\ref{fig:iv_mixed_2}).
As compared to the $V_{\rm g}=-U/2$ case, finite-size effects are somewhat more significant, as in our set-up, the average filling 
of the leads slightly exceeds hall-filling whenever $V_{\rm g}\not= -U/2$. The figure further includes the respective fRG results,\cite{jakobs09}
with a very good agreement with our extrapolated data. Slight deviations between fRG and tDMRG 
at intermediate bias voltages can be attributed to: (i) uncertainties in the extrapolation of the finite-size data and (ii)
the curvature of the bands.
Note  that   band effects
already yield small differences in the  linear conductance as obtained from tDMRG  when compared to results valid in the
wide-band (linear dispersion) limit (see, {\it e.g.}, Fig.~13 in Ref.~\onlinecite{alhassanieh06}).

\section{Summary}
\label{sec:sum}

In this work, we carried out large scale and accurate time-dependent simulations using the 
adaptive tDMRG method to study transport in the single-impurity Anderson model at large biases. 
By performing a finite-size analysis with chains of up to $N=128$ sites, we were able 
to compute the current-voltage characteristics at the particle-hole symmetric point 
up to $U/\Gamma=8$. Convergence with system size is very fast in the voltage regime 
$T_{\rm K}\lesssim V \leq 2t $,
while the resolution in the Kondo regime, {\it i.e.}, $V\lesssim T_{\rm K}$, 
is hampered by finite-size effects, due to the 
exponentially large Kondo correlations\cite{hewson} (see Ref.~\onlinecite{dasilva08},
though).  Our 
results are in excellent agreement with 
those from  fRG\cite{jakobs09} for $U/\Gamma \leq 8$, 
lending strong support to the validity of both approaches.
Where converged, our data thus provide an unbiased benchmark 
from a quasi-exact method for analytical or approximate numerical approaches. 
Furthermore, we studied the mixed-valence regime 
by varying the gate potential at  fixed $U/\Gamma$, presenting results for
current-voltage characteristics.

Let us conclude by comparing our numerical approach, the tDMRG, to other
numerical methods recently applied to non-equilibrium transport in the
single-impurity model, namely, real-time QMC simulations, and the
real-time NRG.
 A direct comparison of our results with the QMC data of Ref.~\onlinecite{werner09} 
(not shown here)
reveals a reasonably good agreement at large bias voltages $V>T_K$. At small bias voltages, the QMC data are consistently below
both tDMRG and fRG.  This  can be traced back to the fact that, due to the sign problem,  
QMC simulations 
are limited to short time scales with a maximum time on the
order of
  $3 /\Gamma$,\cite{werner09}
 while in our case, we are able to reliably simulate times on the order of $\tau \lesssim 6/\Gamma$  at large biases $V\sim t$
and $\tau \lesssim 12/\Gamma$ at small biases $V\sim 0.2t$  (compare Fig.~\ref{fig:crr}). We should emphasize, though, that the QMC simulations
start from the dot being decoupled from the leads in the initial state, which is different from our initial condition and which may lead to
a transient behavior different from our case.
An advantage of the real-time QMC\cite{werner09,schiro09} is that it is designed for  the thermodynamic limit, yet we have shown here that a   finite-size analysis
of tDMRG yields very accurate results. Within both methods, the transient regime, {\it i.e.}, the time window between the initial state and the steady-state
regime can be studied.\cite{schmidt08}

Time-dependent NRG results were published for
the differential conductance $\delta G$ 
and large ratios of    $U/\Gamma$ (Ref.~\onlinecite{anders08}). One the one hand, one may generally expect
 time-dependent NRG to be well suited to capture the finite-bias behavior in the 
Kondo regime $V\lesssim T_k$, while on the other hand, the  data presented for $\delta G$ 
 at  discrete bias voltages,
seem to be  somewhat noisy at large  and intermediate biases 
(see Fig.~3c in Ref.~\onlinecite{anders08}), rendering a comparison with our data difficult.  Real-time NRG calculates the steady-state density operator 
and then extracts the current and the nonequilibrium spectral density from that information.  The numerical challenges in obtaining accurate data for steady-state currents  
seem to be (a) keeping a sufficiently large number of states and (b)  artefacts due to the logarithmic discretization.\cite{anders08} 
Real-time NRG simulations can be performed at finite temperatures as well,\cite{anders08} while methods for real-time DMRG simulations at finite temperatures
are under development\cite{verstraete04,feiguin05b,sirker05,barthel09} and may be available for the study of nonequilibrium currents in the near future.

Our work establishes an important application of the adaptive tDMRG to a topic of much interest 
to both experimentalists and theoretical researchers.
As the technique is highly flexible and can easily be adapted to impurity problems beyond 
Eq.~\eqref{eq:ham} such as interacting leads\cite{feiguin08c} or complex geometries,
we are convinced that this tool
will become highly instrumental in future studies
of  nonequilibrium transport in nanoscopic systems.

\acknowledgments

We acknowledge valuable discussions with   N. Andrei, L. Dias da Silva, S. Jakobs, V.
Meden, H. Onishi,  M. Pletyukhov,   G. Roux,
H. Schoeller, and P. Werner. We further thank the authors of Ref.~\onlinecite{jakobs09} for sending us their results for comparison. 
E.D.  was supported in part by the NSF grant No.\ DMR-0706020
and the Division of Materials Science and Engineering, U.S. DOE, under contract
with UT-Battelle, LLC. 
F.H.-M. acknowledges support by the DFG through FOR 912. A.E.F. was supported by  Microsoft Station Q.

\end{document}